\documentclass[reprint,
groupedaddress,
 amsmath,
 amssymb,
 aps,
 prx,
]{revtex4-2}

\usepackage{graphicx}
\usepackage{dcolumn}
\usepackage{bm}
\usepackage{color}
\usepackage{soul}

\begin{document}

\preprint{APS/123-QED}

\title{Total Faraday rotation by the Hall effect in a 2D electron gas}

\author{Vishnunarayanan Suresh}
\affiliation{D\'epartement de Physique and Institut Quantique, Universit\'e de Sherbrooke, Sherbrooke, Qu\'ebec, Canada J1K 2R1
}

\author{Talia J. Martz-Oberlander}
\affiliation{Department of Physics, McGill University, Montr\'eal, Qu\'ebec, Canada H3A 2T8
}

\author{Sujatha Vijayakrishnan}
\affiliation{Department of Physics, McGill University, Montr\'eal, Qu\'ebec, Canada H3A 2T8
}

\author{Loren N. Pfeiffer}
\affiliation{Department of Electrical Engineering, Princeton University, Princeton, NJ 08544, USA}

\author{Ken W. West}
\affiliation{Department of Electrical Engineering, Princeton University, Princeton, NJ 08544, USA}

\author{Guillaume Gervais}
\affiliation{Department of Physics, McGill University, Montr\'eal, Qu\'ebec, Canada H3A 2T8}

\author{Bertrand Reulet}
\affiliation{D\'epartement de Physique and Institut Quantique, Universit\'e de Sherbrooke, Sherbrooke, Qu\'ebec, Canada J1K 2R1}

\author{Thomas Szkopek}
\affiliation{Department of Electrical and Computer Engineering, McGill University, Montr\'eal, Qu\'ebec, Canada H3A 0E9}


\date{\today}

\begin{abstract}

We report the realization of near total Faraday rotation of $\theta_F=1.43$~rad (82$^\circ$) on a single pass through a 2D electron gas (2DEG), approaching the ideal limit of $\pi/2$ rad (90$^\circ$). The corresponding Verdet constant $V = 9.5\times10^{8}~\mathrm{rad}~\mathrm{T}^{-1}~\mathrm{m}^{-1}$, exceeds by approximately one order of magnitude that reported in other material systems. Our measurements were conducted at microwave frequencies ($f=9.2-11.2$~GHz) in a 2DEG with a high dc mobility $\mu = 7\times10^6$ cm$^2$V$^{-1}$s$^{-1}$, in a hollow waveguide at low-magnetic field ($B < 200$~mT). Near-total Faraday rotation is attributed to the Hall effect with weak radiative coupling to the 2DEG in the inertial, collisionless regime, $\omega \tau \gg 1$, where $\tau$ is the charge transport scattering time. A conducting iris was used to realize weak radiative coupling. Under these conditions, Faraday rotation is strongly enhanced away from the dissipation peak at cyclotron resonance. Our work demonstrates that the classical Hall effect could be ideally suited for the implementation of ideal non-reciprocal devices.
\end{abstract}

\maketitle

\section{Introduction}
Faraday rotation of electromagnetic wave polarization by a two-dimensional electron gas (2DEG) is a manifestation of time reversal symmetry breaking, and has thus been used to probe topological properties of low-dimensional systems \cite{Shimano10,crassee2011giant, Shimano13,Pimenov16,Armitage16, Molenkamp17,nedoliuk2019colossal}. In the quantum limit, the quantization of Faraday rotation by a renormalized fine structure constant has been observed \cite{suresh2020quantitative}, following the seminal work of Volkov and coworkers \cite{Volkov85,Volkov86}. The modification of Faraday rotation by the local electromagnetic environment has also been investigated, including super-radiance \cite{zhang2014superradiant,muravev2016fine} and cavity confinement \cite{arakawa2022microwave}. In contrast, Faraday rotation induced by the classical Hall effect in 2DEGs has received comparatively little attention, with the exception of \cite{sounas2013faraday,skulason2015field}, owing to the common expectation that dissipation associated with 2DEG conduction will limit the achievable Faraday rotation angle. Nonetheless, Faraday rotation induced by the classical Hall effect of free charge carriers, as well as interband transitions, have long been studied in semiconductors \cite{cardona2007faraday}.

Here, we report experiments on Faraday rotation of microwave polarization by a high mobility 2DEG loaded in a hollow metal waveguide with a conductive iris. Contrary to expectation, we demonstrate that the classical Hall effect can be used to achieve near-total Faraday rotation with minimal Ohmic dissipation. A Faraday rotation of $\theta_F=1.43$~rad (82$^\circ$) at $f$ = 10.2 GHz on a single pass through the 2DEG was observed, approaching the ideal limit of $\pi/2$ rad (90$^\circ$). The inertial collisionless regime, where 2DEG displacement current dominates over Ohmic conduction current, corresponds to $\omega\tau \gg$ 1, where $\tau$ is the mean charge transport scattering time and $\omega=2\pi f$. In our experiments, $\omega\tau>17$ at the conditions corresponding to largest Faraday rotation.


The Faraday rotation of a medium is typically quantified by the Verdet constant, $V = \theta_F B^{-1}d^{-1}$, where $B$ is the externally applied magnetic field and $d$ is the path length through the medium. Our experimental result, with $B = 30~\mathrm{mT}$ and GaAs quantum well thickness $d=50~\mathrm{nm}$ corresponds to a Verdet constant of $V = 9.5\times10^{8}~\mathrm{rad}~\mathrm{T}^{-1}~\mathrm{m}^{-1}$, exceeding by approximately one order of magnitude that of monolayer graphene in the THz domain ($1.1\times10^{8}~\mathrm{rad}~\mathrm{T}^{-1}~\mathrm{m}^{-1}$) \cite{nedoliuk2019colossal}, and that of semiconducting monolayer WSe$_2$ in the optical domain ($3.3\times10^{7}~\mathrm{rad}~\mathrm{T}^{-1}~\mathrm{m}^{-1}$) \cite{carey2024giant}. Similarly, our reported Verdet constant exceeds by 4 orders of magnitude that of rare-earth nanoparticle composites in the optical domain ($1\times10^{4}~\mathrm{rad}~\mathrm{T}^{-1}~\mathrm{m}^{-1}$) \cite{bartos2024high}, as well as that of state-of-the-art slow-light atomic vapour cells in the microwave domain ($3\times10^{4}~\mathrm{rad}~\mathrm{T}^{-1}~\mathrm{m}^{-1}$) \cite{siddons2009gigahertz}.

Finally, our experimental results are in excellent agreement with a theoretical analysis based on a classical Drude model for the 2DEG conductivity tensor $\sigma$ at frequency $\omega$, 
\begin{equation}
    \sigma = \frac{\sigma_0}{ \left(1-i\omega \tau \right)^2+\left( \omega_c \tau \right)^2 } \left(\begin{array}{cc} 1 - i \omega \tau & -\omega_c \tau \\ +\omega_c \tau & 1 - i \omega \tau \end{array}\right)
    \label{eq:Drude}
\end{equation}
where $\tau$ is the mean transport scattering time, $\omega_c = e B / m^*$ is the cyclotron frequency, and $\sigma_0 = e n \mu$ is the Drude conductivity of the 2DEG with effective mass $m^*=0.067~m_0$, electron density $n$ and mobility $\mu=e \tau/m^*$. With a simple Drude model, we show that near total Faraday rotation is achieved by the Hall effect with weak, reactive, radiative coupling to a 2DEG in the inertial, collisionless regime.

\section{Experimental methods}

\begin{figure}
    \centering
    \includegraphics[width=1.0\linewidth]{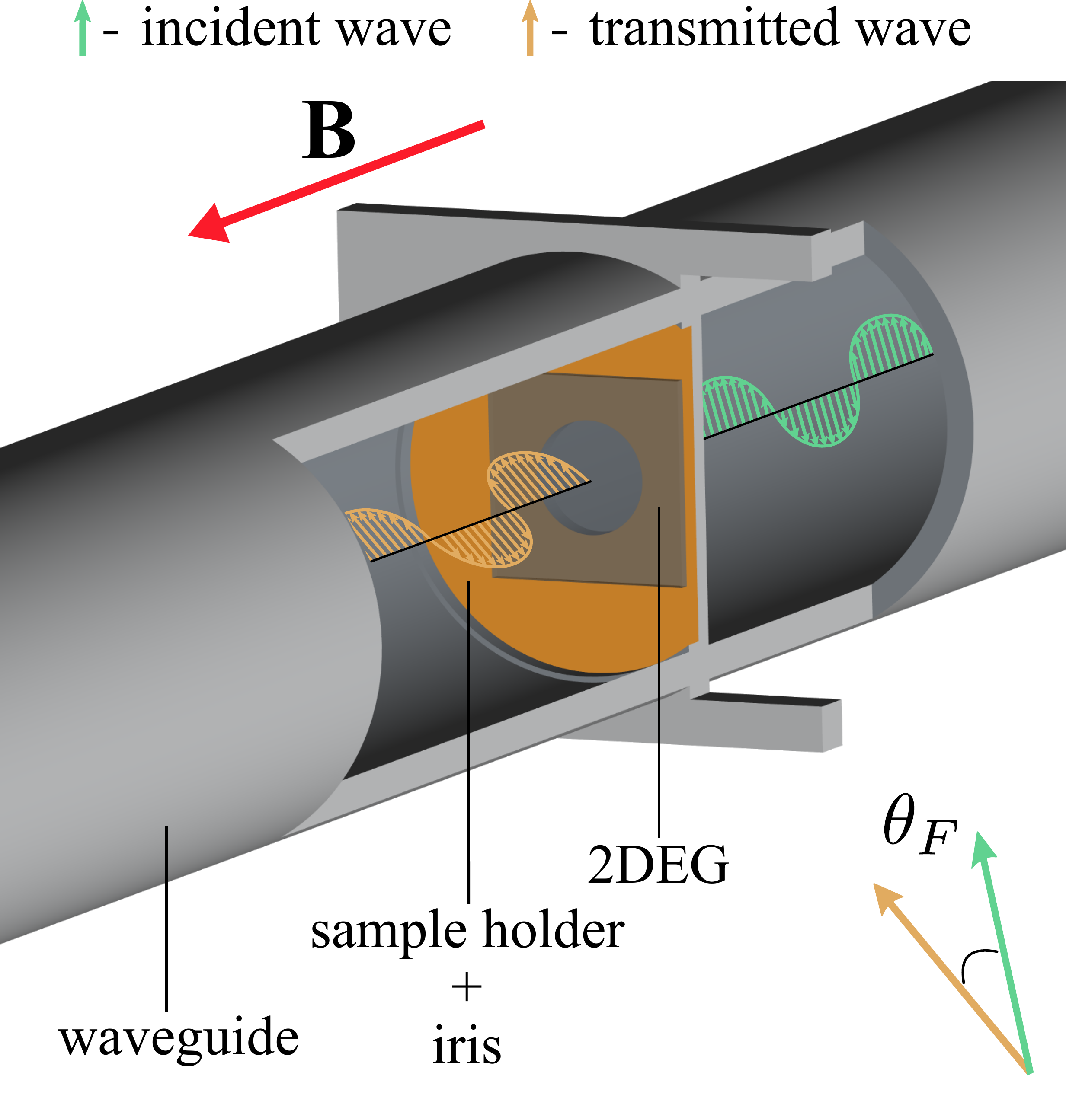}
    \caption{Schematic of the experiment: interior cut-away view of the waveguide with sample holder (iris) and 2DEG. The direction of the applied magnetic field is indicated by the red arrow. The incident and transmitted waves are shown in light green and light orange colors respectively.}
    \label{fig:setup}
\end{figure}

The geometry of our experiment is shown in Fig. \ref{fig:setup}, wherein a 2DEG hosted in a modulation-doped GaAs/AlGaAs quantum well on a GaAs substrate was loaded inside a conducting iris within a hollow cylindrical waveguide. The electron density and mobility of the 2DEG were found to be $n=1.1\times10^{11}~\mathrm{cm}^{-2}$ and $\mu=7\times10^{6}~\mathrm{cm}^{2}~\mathrm{V}^{-1}~\mathrm{s}^{-1}$, respectively, by quasi-dc transport measurements in van der Pauw geometry at $T=10$~mK. The semiconductor sample is a symmetrically doped AlGaAs/GaAs/AlGaAs quantum well of thickness d = 50 nm with a donor setback distance of approximately 256 nm. The hollow cylindrical guide of 23.825~mm  diameter (7.38~GHz cut-off frequency) was silver-plated to reduce loss, and hosts a pair of degenerate, orthogonally polarized TE$_{11}$ modes, which form the basis for polarized excitation and measurement. The 1~cm~$\times$~1~cm GaAs substrate was fitted into a square recess on a metal plate with vacuum grease. A 9~mm diameter circular aperture in the metal plate, co-axial with the waveguide, serves as an iris to couple to the electromagnetic field of the modes near the centre of the waveguide.

The entire assembly was loaded into a cryostat with a superconducting solenoid to generate magnetic field coaxial with the hollow waveguide. All high-frequency measurements were performed at $T = 4$~K. The metal plate with the GaAs substrate and aperture was thermally anchored to  refrigerator with a silver braid. Linearly polarized microwave excitation was injected into the assembly through a linear dipole coupler, and polarization resolved transmission was measured with orthogonally oriented dipole couplers. This assembly is similar to that employed in a previous study of the quantization of Faraday rotation angle \cite{suresh2020quantitative}.

\section{Results and Discussion}

\subsection{Microwave Transmission and Faraday Rotation}

\begin{figure}
    \centering
    \includegraphics[width=\linewidth]{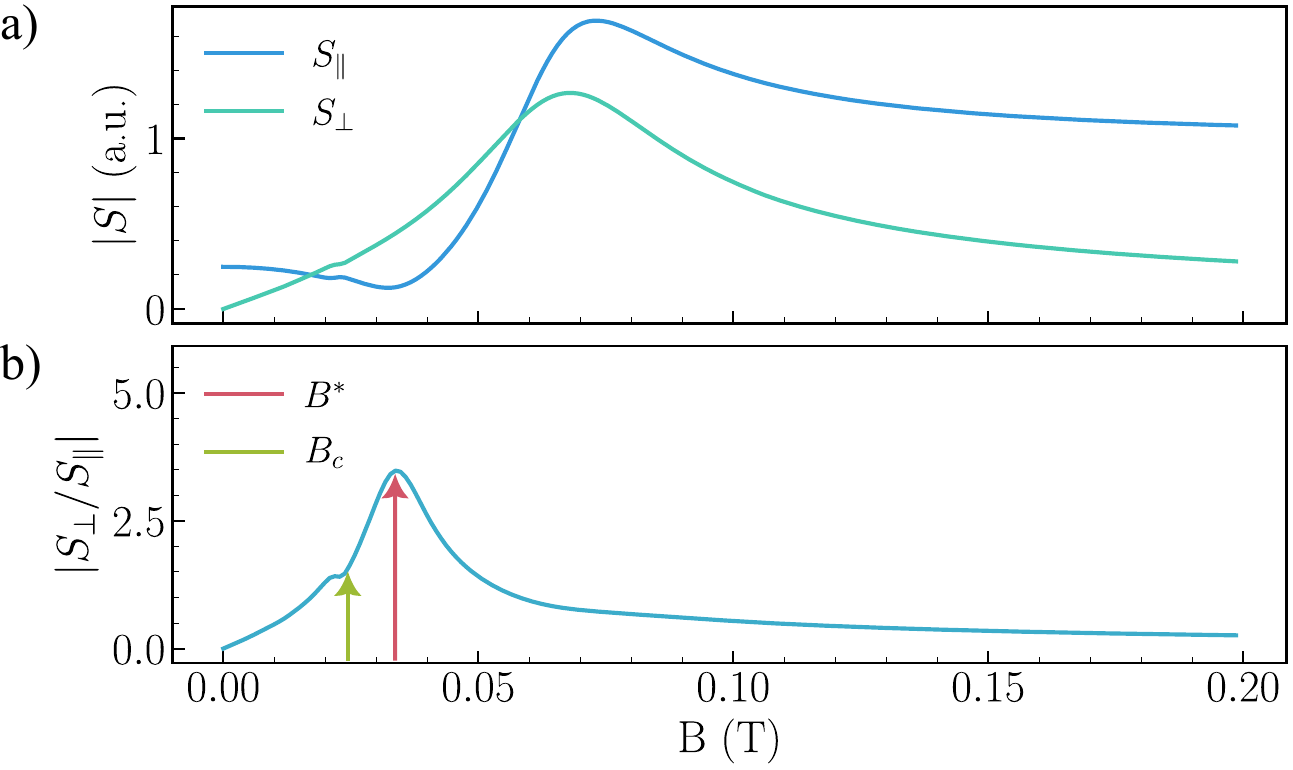}
    \caption{(a) Magnitude of the parallel and perpendicular transmissions, $|S_{\parallel}|$ and $|S_{\perp}|$ vs. magnetic field at frequency 10.2~GHz. (b) $|\tan \left( \theta_F \right)|$ vs. magnetic field at the same frequency. Arrows indicate the fields that correspond to cyclotron resonance, $B_c$ and maximum Faraday rotation, $B^*$}
    \label{fig:transmission}
\end{figure}

The magnitude of the microwave transmission coefficients through the 2DEG versus magnetic field $B$ at a representative frequency $f=10.2$~GHz are shown in Fig. \ref{fig:transmission}(a). The coefficients $S_{\parallel}$ and $S_{\perp}$ correspond to parallel (aligned) and perpendicular (crossed) polarization, respectively. Details concerning the determination of these coefficients from measured scattering parameters are presented in the appendix \ref{sec:analysis}. At $B=0$, the 2DEG is a good conductor with high longitudinal conductivity, and thus high reflectivity, resulting in small parallel transmission $|S_{\parallel}|$. Furthermore, $|S_{\perp}|=0$ because the transverse conductivity is zero in the absence of magnetic field. The non-monotonic magnetic field dependence of $|S_{\parallel}|$ and $|S_{\perp}|$ is determined by the field dependent conductivity tensor.

The transmission coefficients determine the tangent of the Faraday rotation angle $\theta_F$ by
\begin{equation}
    \tan \left( \theta_F \right) = S_{\perp}/S_{\parallel}.
\end{equation}
The magnitude of the Faraday rotation tangent $|\tan \left( \theta_F \right)|$ is shown in Fig. \ref{fig:transmission}(b) versus $B$ at $f=10.2$~GHz. The magnetic field $B_c = \omega(e/m^*)^{-1}$ corresponding to cyclotron resonance and the magnetic field $B^*$ corresponding to maximum Faraday rotation are both indicated. Importantly, the magnetic field corresponding to maximum Faraday rotation is significantly detuned from cyclotron resonance, $B^* > B_c$.

\begin{figure}
    \centering
    \includegraphics[width=\linewidth]{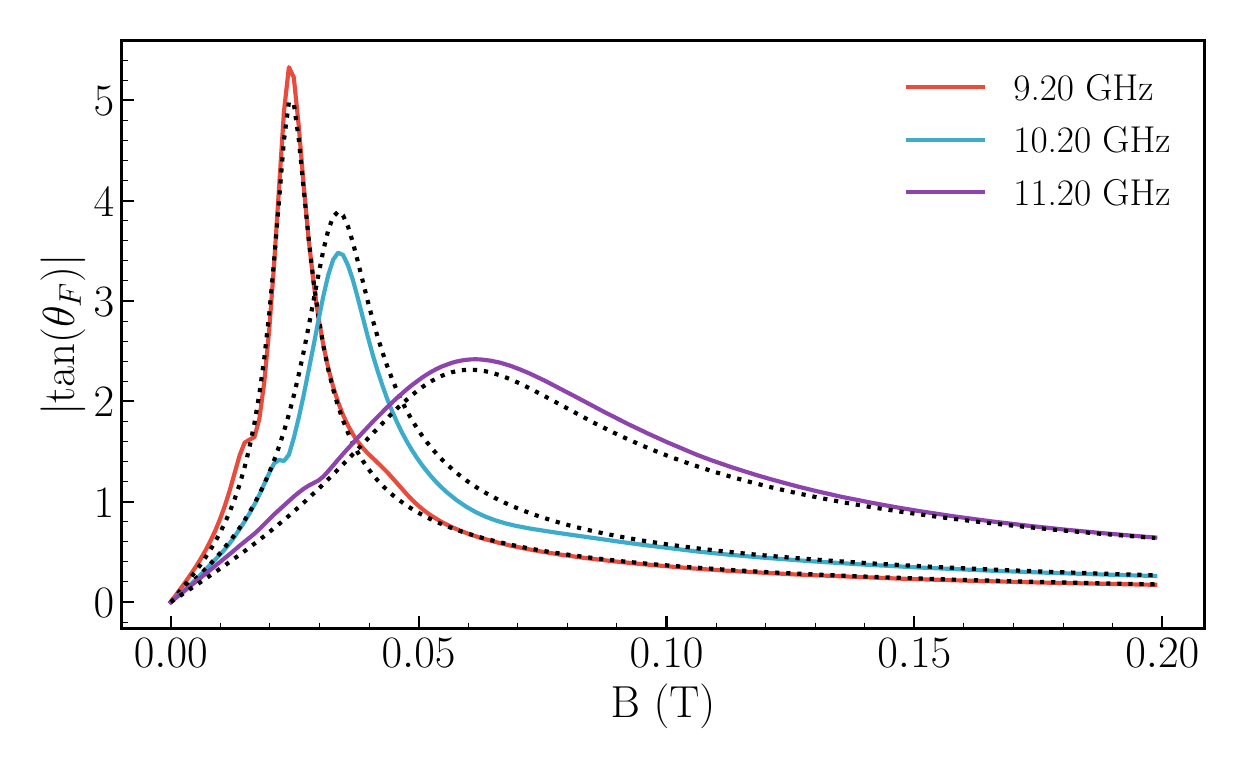}
    \caption{Magnitude of the Faraday rotation tangent $|\tan(\theta_F)|$ vs. magnetic field for different frequencies. Solid lines are data, and dashed lines are theoretical fits (see text). The amplitude of the theoretical curves have been scaled to match experimental data. }
    \label{fig:tan_theta_theory}
\end{figure}

The detuning $B^* > B_c$ is a consequence of the electromagnetic environment of the 2DEG, as will be shown further below. The experimentally measured magnitude of Faraday rotation tangent $| \tan \left( \theta_F \right)|$ versus magnetic field $B$ for frequencies $f=9.20$, $10.20$ and $11.20$~GHz is shown in Fig. \ref{fig:tan_theta_theory}. These measurements reveal frequency dependent variation in: i) the maximum rotation angle $\theta_F$, ii) the magnetic field $B^*$ at which maximum rotation occurs, and iii) the shape of $|\tan \left( \theta_F \right)|$ versus $B$. Theoretical model fits are also shown in Fig. \ref{fig:tan_theta_theory} in dashed lines, revealing excellent agreement over the entire range of frequencies and field explored in this work.

We apply here a theoretical model for Faraday rotation based on the linear transport \textit{ansatz} of \cite{suresh2020quantitative}, consisting of the general relation,
\begin{equation}
\tan(\theta_F) = \frac{ \gamma Z \sigma_{yx} }{K + Z \sigma_{xx}},
\end{equation}
where $Z$ is a wave impedance, $K$ is a transmission coefficient, and $\gamma$ is a mode-coupling parameter. Substitution of the classical Drude conductivity of Eq.(\ref{eq:Drude}) gives,
\begin{equation}
\tan(\theta_F) = \frac{\alpha}{\tilde \mu} \frac{ B }{ B^2 - B_0^2 },
\label{eq:tan_teta}
\end{equation}
where $\alpha = (Z/K) \tilde\sigma_0$ is a normalized coupling constant between the 2DEG and waveguide modes, $\tilde\sigma_0= \sigma_0 ( 1 - i \omega \tau )^{-1}$ and $\tilde \mu = \mu ( 1 - i \omega \tau )^{-1}$ are generalized conductivity and mobility taking into account inertial delay in electron response, and $B_0 = i(1+\alpha)^{1/2} \tilde \mu^{-1}$ is the complex pole of $\tan(\theta)$. The mode-coupling parameter is taken as $\gamma=1$ since the mode field profile approximates that of a plane wave at the centre of the guide where the 2DEG is located in the iris. Within this model, maximum Faraday rotation occurs at a magnetic field $B^* =|B_0|$, and the width of the peak in Faraday rotation is determined by $\Im \textrm{m}\left\{ B_0 \right\}$. 

\subsection{Generalized Mobility and Electromagnetic Coupling}

We can gain greater insight into the experimental results by consideration of the frequency dependent model parameters $\tilde \mu$, $\alpha$ and $B_0$. To this end, for each frequency $f$, the magnetic field dependent $\tan(\theta_F)$ that was measured was fit to the model of Eq. (\ref{eq:tan_teta}) to determine $\tilde \mu$, $\alpha$ and $B_0$. The details of the fit procedure are provided in the appendix \ref{sec:analysis}.

\begin{figure}
    \centering
    \includegraphics[width=\linewidth]{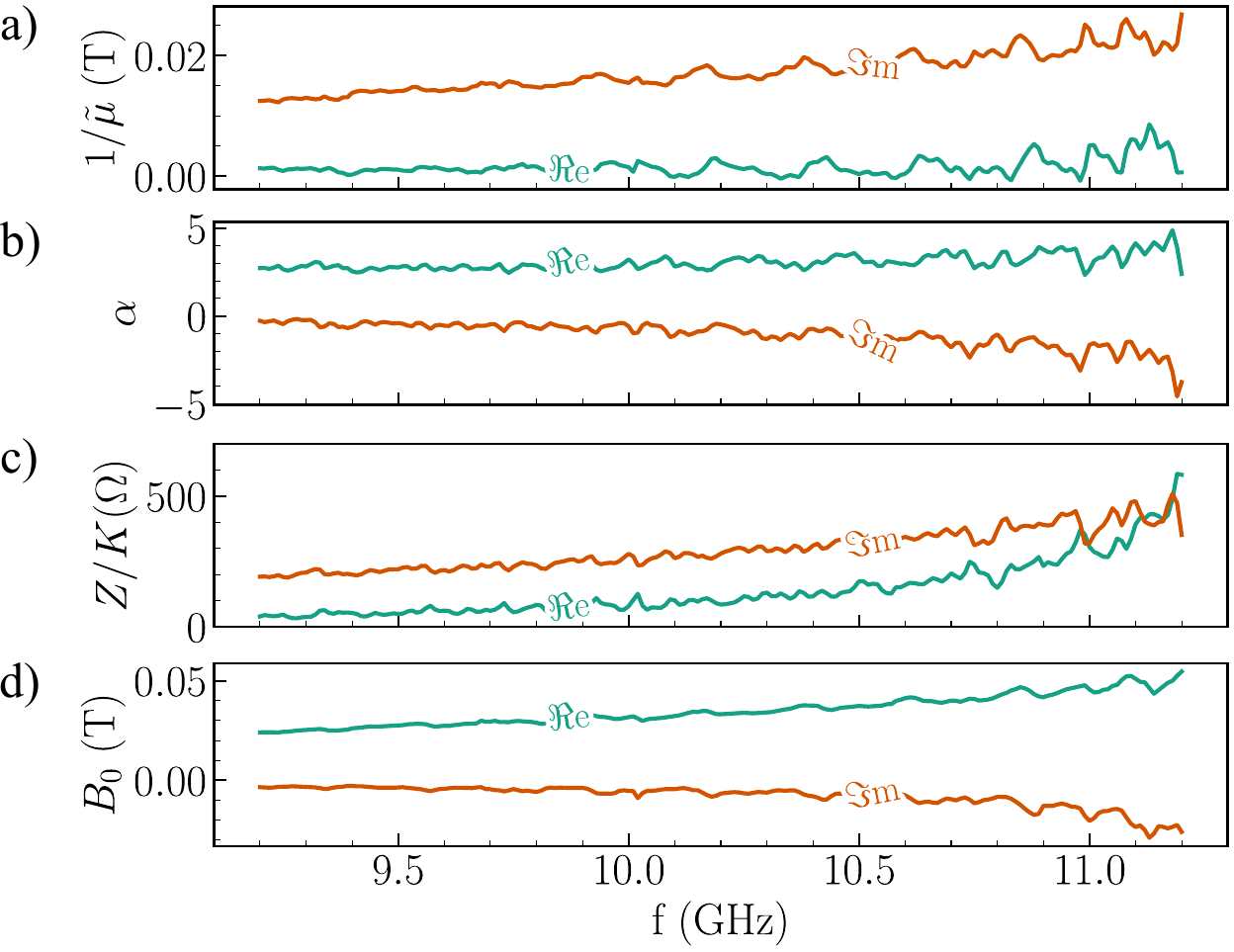}
    \caption{(a) Reciprocal mobility $\tilde\mu^{-1}$ versus frequency $f$, (b) electromagnetic coupling parameter $\alpha$ versus frequency $f$. (c) impedance parameter $Z/K$ versus frequency $f$, and (d) Faraday rotation pole $B_0$ versus frequency $f$. Each quantity is resolved into real and imaginary components. }
    \label{fig:model_params}
\end{figure}

The reciprocal of the generalized mobility $\tilde \mu^{-1}$ is shown versus frequency in Fig. \ref{fig:model_params}a, where it is seen that $\tilde \mu^{-1}$ is predominantly imaginary, $\Im\textrm{m} \left\{ \tilde \mu^{-1} \right\} \gg \Re\textrm{e} \left\{ \tilde \mu^{-1} \right\}$. This implies that the high mobility 2DEG is in the inertial, collisionless limit in the microwave frequency domain with $\omega \tau \gg 1$, since $\tilde \mu^{-1}=( 1-i\omega \tau )\mu^{-1}$ and $\mu$ is real. Indeed, $\Re\textrm{e} \left\{ \tilde \mu^{-1} \right\}$ is sufficiently small that it cannot be reliably determined. The transport scattering time inferred from dc transport measurements is $\tau = \mu_{dc} (m^*/e) = 0.26~\mathrm{ns}$, from which it follows $\omega \tau = 17$ at $f = 10.2$~GHz, consistent with our Faraday rotation model fit.

In the inertial, collisionless limit, the Drude conductivity reads:
\begin{equation}
    \lim_{\omega \tau \gg 1} \sigma = \frac{\sigma_0}{ \left(\omega_c^2 - \omega^2 \right)  \tau^2 } \left(\begin{array}{cc} -i \omega \tau & -\omega_c \tau \\ +\omega_c \tau & -i \omega \tau \end{array}\right),
    \label{eq:Drude_coll}
\end{equation}
with a purely reactive longitudinal conductivity. Expressed in a circular polarization basis,
\begin{equation}
    \lim_{\omega \tau \gg 1} \sigma_{\pm} = \sigma_{xx} \pm i\sigma_{xy} = \frac{\sigma_0}{ i(\omega \pm \omega_c)\tau},
\end{equation}
where the absence of damping implies negligible Ohmic dissipation in the collisionless limit. Note that frequency dependent ripple can be observed in $\tilde \mu^{-1}$ in Fig. \ref{fig:model_params}a, which we attribute to remnant standing wave effects originating from imperfect calibration of the entire microwave assembly.

Electromagnetic coupling with the 2DEG is quantified by the dimensionless coupling parameter $\alpha$, as shown versus frequency in Fig. \ref{fig:model_params}b. It can be observed that $\alpha$ is predominantly real within the lower frequency range of our experiments. Using $\omega \tau = 17$ determined above, $Z/K=\alpha/\tilde\sigma_0$ as shown in Fig. \ref{fig:model_params}c. At lower frequencies, the impedance $Z/K$ is dominated by a positive imaginary component, corresponding to a capacitive reactive coupling (with $\exp(-i\omega t)$ convention) between the 2DEG current and the local electromagnetic environment. A 2DEG in free-space exhibits a real $Z/K$, corresponding to ideal radiation resistance \cite{suresh2020quantitative}. Here, the reactive coupling of the 2DEG current to propagating waveguide modes arises from the mode propagation cut-off of the iris. The $d=9$~mm diameter of the iris has a cut-off frequency $f_0 = \chi'_{1,1}(\pi c d)^{-1}=19.4$~GHz, where $c$ is the speed of light and $\chi'_{1,1}=1.8412$ is a Bessel function zero. In other words, the 2DEG current excites significant near-fields dressing the iris, leading to capacitive coupling.

The combination of the collisionless 2DEG regime and the reactive coupling of the 2DEG to the electromagnetic environment leads to an increase in the maximum of $\tan (\theta_F)$. The complex pole $B_0$ in Faraday tangent is shown versus frequency in Fig. \ref{fig:model_params}d. The peak in Faraday rotation tangent diverges as $\Im\textrm{m}\{B_0\} \rightarrow 0$, see Eq.(\ref{eq:tan_teta}). This ideal scenario corresponds to total Faraday rotation, where $\theta_F \rightarrow \pi/2$. As seen in Fig. \ref{fig:model_params}d, there is a vanishing small $\Im\textrm{m}\{B_0\}$ in the low frequency regime, as a consequence of an imaginary $\tilde \mu^{-1}$ from the collisionless 2DEG regime, and real $\alpha$ from reactive coupling. The broadening in $\tan (\theta_F)$ versus $B$ as frequency $f$ increases, observed in Fig. \ref{fig:tan_theta_theory}, corresponds to the increase in $\Im\textrm{m}\{B_0\}$ as frequency increases.

The detuning between the magnetic field of peak Faraday rotation $B^*=|B_0|$ and the cyclotron resonance field $B_c=\omega (e/m^*)^{-1}$ is determined by the strength of electromagnetic coupling in the collisionless limit,
\begin{equation}
    \lim_{\omega \tau \gg 1} |B_0| =|1+\alpha|^{1/2}B_c.
\end{equation}
Thus, increasing the magnitude of electromagnetic coupling $\alpha$ increases the detuning of peak rotation $B^*$ from cyclotron resonance $B_c$ in the collisionless limit.

Interestingly, in the common scenario of dissipative conduction, $\omega \tau \ll 1$, and weak electromagnetic coupling, $|\alpha| \ll 1$,
\begin{equation}
    \lim_{\omega \tau \ll 1} B_0 = \mu^{-1},
\end{equation}
revealing the mobility $\mu$ as the sole factor determining the field of peak Faraday rotation. In this dissipative, weakly-coupled regime, the peak rotation tangent is $\max \{ \tan (\theta_F) \} = \alpha/2 = Z \sigma_0/(2K)$. This is in stark contrast to the dissipationless, reactively coupled regime, where $\tan(\theta_F)$ diverges, and total rotation is approached, $\theta_F \rightarrow \pi/2$.

\subsection{Polarization Analysis and Total Rotation}

\begin{figure}
    \centering
    \includegraphics[width=\linewidth]%
    {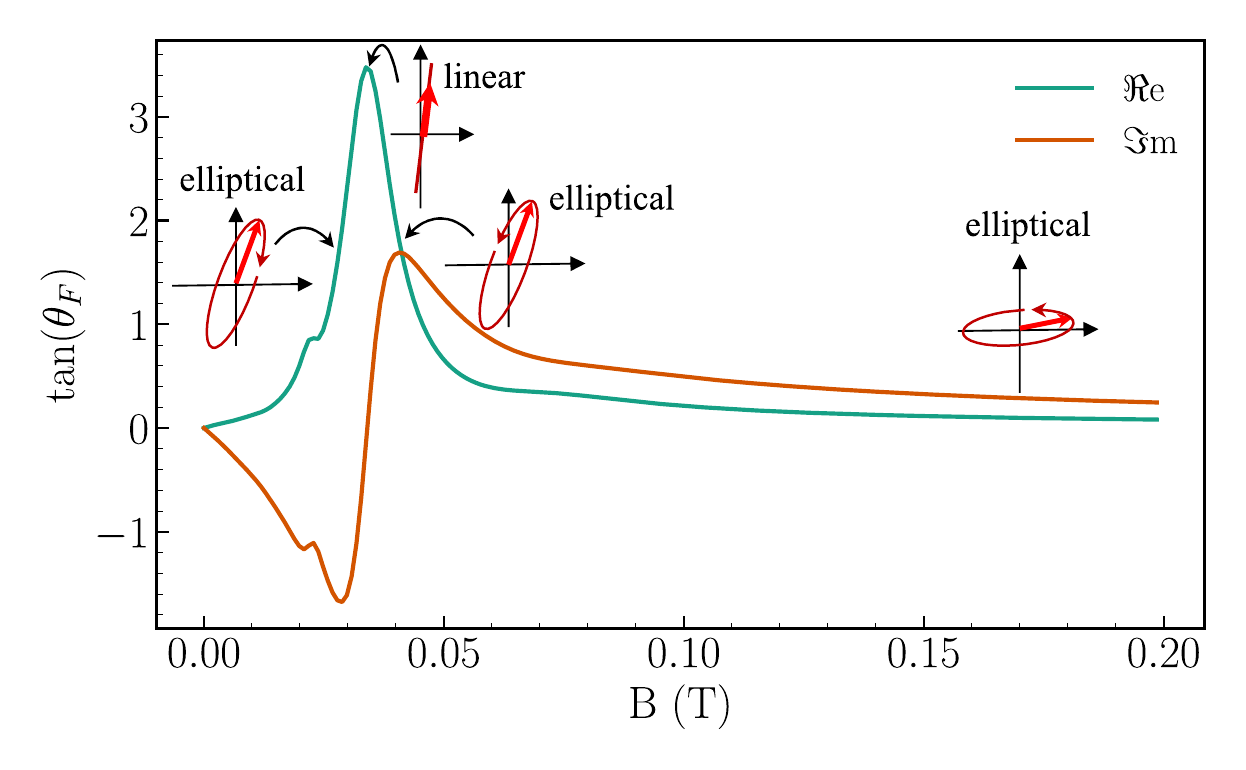}
    \caption{ The real and imaginary components of $\tan(\theta_F)$ versus magnetic field $B$ at $10.2$~GHz. The polarization ellipse at representative magnetic fields are also shown.}
    \label{fig:pol}
\end{figure}

The polarization state of the field transmitted through the 2DEG is in general elliptical owing to the phase delay $\delta$ between the perpendicular and parallel transmission coefficients: $\tan(\theta_F)=|\tan(\theta_F)| \exp(+i\delta)$. The real and imaginary components of $\tan(\theta_F)$ are shown in Fig. \ref{fig:pol} versus $B$ at $f=10.2$~GHz. Details regarding correction for the phase delay between the parallel and perpendicular transmission paths are discussed in Appendix \ref{sec:analysis}. Representations of the corresponding polarization ellipse at select magnetic fields show the evolution of the polarization state versus magnetic field.

At the field $B^*$ of maximum $|\tan(\theta_F)|$, the imaginary component of $\tan(\theta_F)$ is zero and thus $\delta=0$, corresponding to linear polarization. This is another feature of Faraday rotation by a high mobility 2DEG in the collisionless limit with reactive electromagnetic coupling.

\begin{figure}
    \centering
    \includegraphics[width=\linewidth]
    {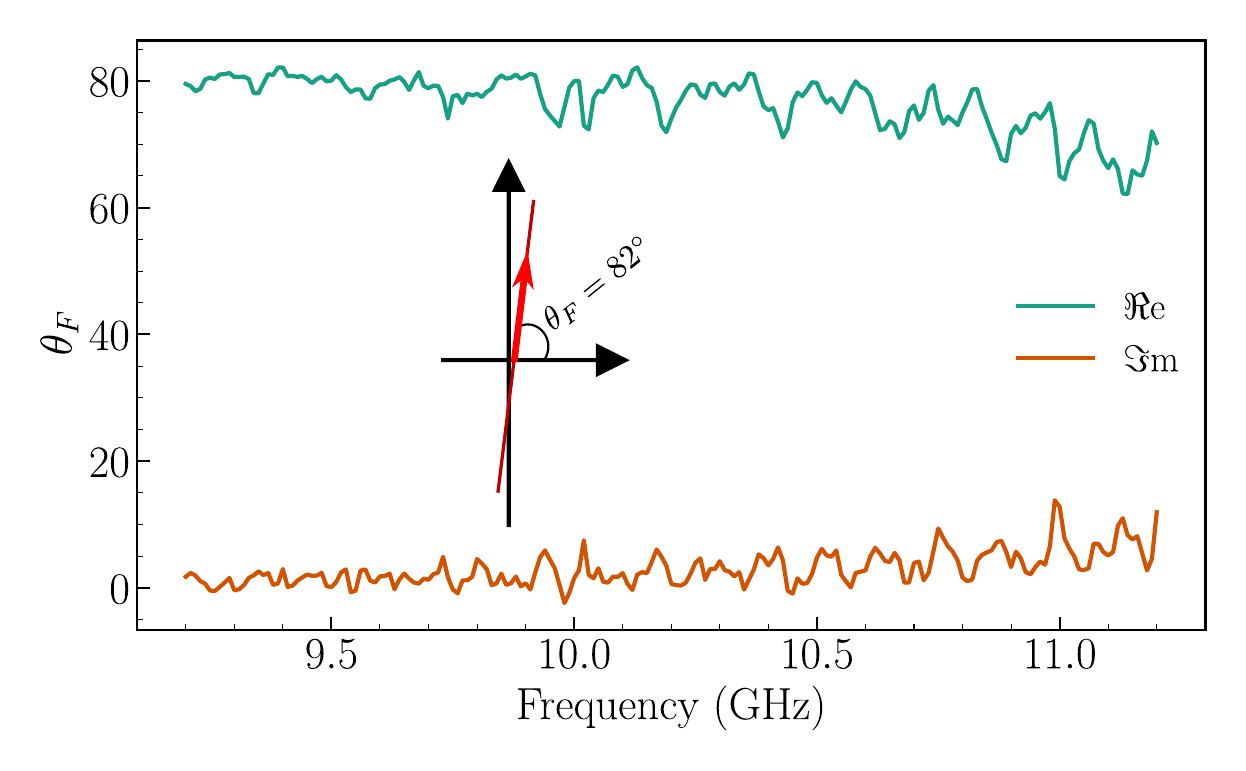}
    \caption{ The real and imaginary components of the maximum  Faraday rotation angle $\theta_F$ vs. frequency $f$. A maximum $\Re\textrm{e}\{\theta_F\}=82^{\circ}$ on a single pass through the 2DEG is observed }
    \label{fig:max_rot}
\end{figure}

The experimentally determined real and imaginary components of maximum $\theta_F$ are shown in Fig. \ref{fig:max_rot} versus frequency $f$, where the complex Faraday rotation angle, $\theta_F = \arctan(\tan(\theta_F)) = \arctan(S_\perp/S_\parallel)$. The real (imaginary) part of $\theta_F$ accounts for the in-phase (out-of-phase) component of the transmission. Notably, a maximum $\Re\textrm{e}\{\theta_F\}=82^{\circ}$ ($1.43$~rad) is achieved, with $\Re\textrm{e}\{\theta_F\}\sim 80^{\circ}$ ($1.40$~rad)  over a broad frequency range. The Faraday rotation on a single pass through the 2DEG is thus found to approach the ideal limit of total Faraday rotation, $\theta_F=90^\circ$ ($\pi/2$).
The small value of $\Im\textrm{m}\{\theta_F\}$ as compared to $\Re\textrm{e}\{\theta_F\}$ corresponds to a linear polarization state (phase delay $\delta \rightarrow 0$).

\section{Conclusions}

In summary, we have shown that the classical Hall effect can produce near total Faraday rotation $\theta_F = \pi/2$, provided charge transport is in the collisionless regime and electromagnetic coupling is reactive. Looking forward, these principles can be applied to material systems beyond 2DEGs hosted in semiconductor heterostructures, and to electromagnetic environments beyond hollow metal waveguides. Nor is the spectral range of operation limited to the microwave domain. Notably, Faraday rotation is the fundamental process at the heart of most non-reciprocal electromagnetic devices, including isolators and circulators. It is foreseeable that more complex geometries employing the near total Faraday rotation by classical Hall effect reported here could be used to realize non-reciprocal devices.

\appendix

\section{Data Analysis\label{sec:analysis}}

For a perfect orthomode transducer and a perfect alignment with the emission antenna, the measured complex transmission coefficients $t_\parallel$ and $t_\perp$ are related to the transmission coefficients $S_\parallel$, $S_\perp$ of the 2DEG by: $t_\parallel =g_\parallel S_\parallel$, $t_\perp= g_\perp S_\perp$, where $g_\parallel$ and $g_\perp$ are gains (and/or losses) particular to the parallel and perpendicular polarizations, respectively. An imperfect alignment between the emission/detection antennas leads to a mixing between the two. Since $S_\parallel(B)$ is an even function of the magnetic field while $S_\perp(B)$ is odd, the mixing leads to $t_\parallel$ being not perfectly even and $t_\perp$ not perfectly odd. We found these effects to be small, implying that the misalignment is small. Moreover, considering the even part of $t_\parallel$ and the odd part of $t_\perp$ removes the mixing. In our setup, thanks to the use of a cryogenic switch, the paths followed by the parallel and perpendicular signals differ only by a cable and the port of the orthomode transducer, so we expect $g_\parallel$ and $g_\perp$ to be of comparable amplitude and with a phase difference that corresponds to a delay. Yet both are frequency-dependent.

In the following we will consider separately the two transmission coefficients, related to the conductivity of the 2DEG by,
\begin{eqnarray}
    S_\parallel &=& \frac{K+Z\sigma_{xx}}{\Delta},\\
    S_\perp &=& \frac{\gamma Z \sigma_{yx}}{\Delta},\\
    \Delta &=& (K+Z\sigma_{xx})^2+(\gamma Z\sigma_{yx})^2.
\end{eqnarray}
We introduce the renormalized conductivity $\tilde\sigma_0=\sigma_0/(1-i\omega\tau)$ and renormalized mobility $\tilde\mu=\mu/(1-i\omega\tau)$ to write the Drude conductivity tensor as
\begin{eqnarray}
    \sigma_{xx} &=& \tilde\sigma_0 \frac{1}{1+(\tilde\mu B)^2}, \\
    \sigma_{yx} &=& \tilde\mu B \sigma_{xx}.
\end{eqnarray}

We determine the parameter $\alpha=Z\tilde\sigma_0/K$, from the ratio of $S_\parallel$ at high and low field
\begin{equation}
    \frac{S_\parallel(B\rightarrow\infty)}{S_\parallel(B=0)}=\frac{t_\parallel(B\rightarrow\infty)}{t_\parallel(B=0)}=1+\alpha,
\end{equation}

and we then calculate the ratio $z=t_\perp/t_\parallel\propto\tan(\theta)$ by
\begin{equation}
    z=a\frac {BB_0}{B^2+B_0^2},
    \label{eq:z}
\end{equation}
with
\begin{eqnarray}
    (\tilde\mu B_0)^2 &=& 1+\alpha \label{eq:mu_tilde},\\
    a &=& \frac{g_\perp}{g_\parallel}\frac{\alpha}{\sqrt{1+\alpha}}\gamma. \label{eq:a}
\end{eqnarray}

Fitting the field dependence of $|z(B)|$ using Eq.(\ref{eq:z}) provides a precise determination of the complex constant $B_0$, (see Fig. \ref{fig:tan_theta_theory} for examples of fits). $B_0$ is plotted in Fig. \ref{fig:model_params}(d) vs. field.

From $\alpha$ and $B_0$ we can in principle deduce $\tilde\mu$ using Eq.(\ref{eq:mu_tilde}). We show in Fig. \ref{fig:model_params}(a) the real and imaginary parts of $1/\tilde\mu=(1-i\omega\tau)/\mu$. We observe that $\Re\textrm{e} (1/\tilde\mu)$ oscillates around $10^{-3}$, sometimes becoming negative. This is consistent with the very high value of the dc mobility. From our high frequency analysis, we can only conclude that $\mu\sim 1000~\textrm{T}^{-1}$ ($10^7\textrm{cm}^{-2} \textrm{V}^{-1} \textrm{s}^{-1}$). An estimate for the value of $\tau$ is thus obtained from the dc transport relation, $\mu=\mu_{dc}=e\tau/m^*=700~\textrm{T}^{-1}$. From $\alpha$ we calculate $Z/K=\alpha/\tilde\sigma_0$ using $\tilde\sigma_0=ne\tilde\mu$, as shown in Fig.~\ref{fig:model_params}(c).


We then separately consider $S_\parallel$ and $S_\perp$, not only their ratio. Focusing on the very low field part, we observe a small resonance below $B_0$, which corresponds to the cyclotron resonance. Attempts to fit this resonance showed its strong sensitivity to $\gamma$, which has to be chosen close to $1$. With this value, the overall amplitudes of $S_\parallel$ and $S_\perp$ allow us to deduce the gains $g_\parallel$ and $g_\perp$.

\section*{Acknowledgments}

We thank René Côté for fruitful discussions. We are very grateful to Edouard Pinsolle, Gabriel Laliberté and Christian Lupien for their technical help. GG, BR and TS thank the Fonds de Recherche du Qu\'ebec, Nature et Technologies (FRQNT) and the Natural Sciences and Engineering Research Council of Canada (NSERC) for financial support. The Princeton University portion of this research was funded in part by the Gordon and Betty Moore Foundation’s EPiQS Initiative, Grant GBMF9615.01 to LP. BR acknowledges support from the Canada Research Chair program, the Canada First Research Excellence Fund and the Canada Foundation for Innovation.

\bibliography{Faraday}

\end{document}